# Kinetic Modeling Analysis of Ar Addition to Atmospheric Pressure $N_2$-$H_2$ Plasma for Plasma-Assisted Catalytic Synthesis of $NH_3$

Zihan Lin[a], Shota Abe[a,b], Zhe Chen[a], Surabhi Jaiswal[a], and Bruce E. Koel[a,*]

[a]Department of Chemical and Biological Engineering, Princeton University, Princeton, NJ 08544, USA

[b]Princeton Plasma Physics Laboratory, 100 Stellarator Road, Princeton, NJ 08540, USA

[*]Corresponding author: bkoel@princeton.edu

**Abstract**

Zero-dimensional kinetic modeling of atmospheric pressure Ar-$N_2$-$H_2$ nonthermal plasma was carried out to gain mechanistic insights into ammonia formation during plasma-assisted catalysis of ammonia synthesis. The kinetic model was developed for a coaxial dielectric barrier discharge (DBD) quartz wool-packed bed reactor operating at near room temperature using a kHz-frequency plasma source. With 30% Ar mixed in a 1:1 $N_2$-$H_2$ plasma at 760 Torr, we find that $NH_3$ production is dominated by Eley-Rideal (E-R) surface reactions, which heavily involve surface $NH_x$ species derived from N and H radicals in the gas phase, while the influence of excited $N_2$ molecules is negligible. This is contrary to the commonly proposed mechanism that excited $N_2$ molecules created by Penning excitation of $N_2$ by Ar (4s) and Ar(4p) plays a significant role in assisting $NH_3$ formation. Our model shows that the enhanced $NH_3$ formation upon Ar dilution is unlikely due to the interactions between Ar and H species, as excited Ar atoms have a weak effect on H radical formation through $H_2$ dissociation compared to electrons. We find that excited Ar atoms contribute to 28% of the N radical production in the gas phase via $N_2$ dissociation, while the rest is dominated by electron-impact dissociation. Furthermore, Ar species play a negligible role in the product $NH_3$ dissociation. $N_2$ conversion sensitivity analyses were carried out for electron density ($n_e$) and reduced electric field (E/N), and contributions from Ar to gas-phase N radical production were quantified. The model can provide guidance on potential reasons for observing enhanced $NH_3$ formation upon Ar dilution in $N_2$-$H_2$ plasmas beyond changes to the discharge characteristics.

**1. Introduction**

Ammonia is an essential chemical that is heavily used in industrial and agricultural applications. The world currently produces 130 million tons of $NH_3$ annually, most of which is used as a precursor for the synthesis of fertilizers.[1] Current industrial processes for synthesizing ammonia are carried out via the traditional Haber-Bosch process, an energy-intensive approach that relies on elevated temperature (400–500 °C) to increase the reaction rate for $NH_3$ production to a desirable value and high pressure (100–200 bar) to increase the $NH_3$ equilibrium concentration. Given the extreme conditions, industrial ammonia synthesis accounts for 1-2% of total world energy consumption while emitting 300 million metric tons of $CO_2$ annually.[2]

Plasma-assisted catalysis using nonthermal plasma is a promising approach to circumvent the high temperature and pressure required for thermal catalysis through plasma-catalyst synergies. A nonthermal plasma discharge can produce highly reactive species in the gas phase (radicals, ions, free electrons, and excited species) that can undergo reactions at ambient temperature. Most of the energy dissipated to drive a discharge is deposited into free electrons, which generate reactive species through excitation, ionization, and dissociation.[3] In recent years, dielectric barrier discharges (DBD) have become ubiquitous in plasma-assisted catalysis studies due to the effective generation of energetic electrons (1-10 eV) and excited species, and geometric flexibility.[4] In ammonia synthesis, emphasis has been placed on overcoming the $N_2$ dissociation barrier, the rate-limiting step in traditional thermal catalysis.[5] To this end, Mehta et al. proposed that a more energetically efficient pathway is attainable via the adsorption of vibrationally excited $N_2$ molecules on catalytic sites, thereby lowering the energy barrier for $N_2$ dissociation.[6] Alternatively, a less energetically favored approach that has been observed is to produce N radicals by direct $N_2$ dissociation in the plasma.[7] Substantial efforts have been devoted to advancing the plasma-assisted catalytic synthesis of ammonia, especially by studying novel catalysts to achieve a higher energy yield.[2,8–16]

Altering plasma discharge properties via noble gas addition is a well-known concept in surface etching and plasma processing applications.[17–19] Argon plasmas are populated with a high concentration of active species such as metastable Ar, which has a high capability for storing excess energy. Due to their long lifetimes, these metastable states may activate other gaseous species via Penning excitation and ionization processes.[20–22] Noble gas addition can also stabilize the plasma discharge and offer better control over plasma properties.[19,23] These concepts have been applied in different types of discharges for plasma-assisted ammonia synthesis. Nakajima and Sekiguchi reported an increased $NH_3$ formation rate upon argon addition in a microwave discharge.[24] Liu et al. found enhanced $N_2$ conversion by diluting an $N_2$-$H_2$ plasma with up to 30% argon in a quartz wool-packed atmospheric pressure DBD reactor.[25] In a completely different application, such discoveries have also motivated the investigation of ammonia formation on tungsten and stainless steel surfaces for fusion energy applications.[26] Proposed mechanisms for the argon-initiated enhancement of $NH_3$ formation in the gas phase are the production of active N species via Penning excitation,

$$\text{Ar}^* + \text{N}_2 \rightarrow \text{Ar} + \text{N}_2^*, \tag{1}$$

and dissociation reactions by excited Ar species,[25]

$$\text{Ar}^* + \text{N}_2 \rightarrow \text{Ar} + \text{N} + \text{N}, \tag{2}$$

$$\text{Ar}^* + \text{H}_2 \rightarrow \text{Ar} + \text{H} + \text{H}. \tag{3}$$

An alternative explanation was offered by Nakajima and Sekiguchi, who attributed their observation to charge transfer between $\text{Ar}^+$ and $\text{N}_2$, which occurs readily due to their similar ionization energies (~15.7 eV).[24,26,27] However, details of the plasma chemistry resulting in ammonia synthesis are not clear in atmospheric pressure Ar-$\text{N}_2$-$\text{H}_2$ discharges, and experimental results have had limited help from theoretical explanations of modeling studies of Ar-$\text{N}_2$ and Ar-$\text{H}_2$ plasma discharges for different applications.[27–30] Extensive kinetic modeling for $\text{N}_2$-$\text{H}_2$ plasma was conducted by Hong et al. to elucidate important mechanisms for plasma-assisted ammonia synthesis.[1] Sode et al. previously modeled low-pressure inductively coupled Ar-$\text{N}_2$-$\text{H}_2$ plasma with only 1% Ar dilution.[31] Yet, no atmospheric pressure Ar-$\text{N}_2$-$\text{H}_2$ plasma modeling of chemical pathways has been investigated to validate proposed mechanisms that facilitate $\text{NH}_3$ formation.

In this article, we report an analysis of the role of argon in the dominant reaction pathway for ammonia synthesis in a DBD flow reactor packed with quartz wool. We adapted the $\text{N}_2$-$\text{H}_2$ nonthermal plasma model from Hong et al.[1] for understanding ammonia synthesis, consistent with two previous studies from our group on plasma-assisted catalytic synthesis of ammonia in a DBD coaxial flow reactor.[15,16] Herein, we expanded the $\text{N}_2$-$\text{H}_2$ nonthermal plasma kinetic model[1] by including Ar plasma chemistry. We compared our calculated $\text{N}_2$ conversions to existing experimental data to benchmark our results. Detailed pathways for $N_2$ conversion and Ar contributions to $NH_3$ production were analyzed for various plasma conditions.

## 2. Methods

We utilized ZDPlasKin, a 0D kinetics solver, to simulate nonthermal Ar-$N_2$-$H_2$ plasma chemistry for $NH_3$ synthesis that allows for reactions both in the gas phase and on solid surfaces.[32] ZDPlasKin is integrated with a built-in software BOLSIG+, a Boltzmann equation solver that calculates electron properties such as the mean electron temperature and the electron energy distribution function, as well as rate constants of reactions involving electrons.[33,34] We modified ZDPlasKin to simulate gas flow in a continuously stirred tank reactor (CSTR).[15] Species density evolutions are represented by the CSTR design equation

$$\frac{dn_i}{dt} = \frac{Q}{V}(n_{i,0} - n_i) + \sum_j R_{ij}, \qquad (4)$$

where $Q$ is the total input gas flow rate, $V$ is the plasma/reactor volume, $n_{i,0}$ and $n_i$ are the inlet and outlet concentrations of species $i$ expressed in number densities respectively, and $R_{ij}$ is the production rate of species $i$ via reaction $j$. Based on the existing $N_2$-$H_2$ kinetic model[1,15], we added Ar species and reactions to our ZDPlasKin CSTR kinetic model, which are listed in Tables 1 and 2, respectively. Ar(meta) and Ar(r) are the 4s excited states. Ar(meta) denotes the $1s^3$ and $1s^5$ states in Paschen notation. Ar(r) denotes the $1s^2$ and $1s^4$ states. $Ar_2^*$ is electronically excited $Ar_2$. Reactions for N and H species, including interactions between vibrationally and electronically excited species, radicals, ions, and surface species, were adapted from Hong et al.[1] as in our previous reports.[15,16] In our discussion of excited state chemistry, the electronically excited N$_2$ states N$_2$($C^3\Pi_u$), N$_2$($B^3\Pi_g$), and N$_2$($A^3\Sigma_u^+$) are abbreviated as N$_2$(C3), N$_2$(B3) and N$_2$(A3), respectively.

**Table 1.** Ar species included in the atmospheric pressure Ar-N$_2$-H$_2$ plasma kinetic model in addition to the N and H species given in ref [1].

| Neutral atom in ground state | Ions | Excited species |
|---|---|---|
| Ar | Ar$^+$, ArH$^+$, ArH$_2^+$, ArH$_3^+$, Ar$_2^+$, Ar$_2$H$^+$ | Ar(meta), Ar(r), Ar(4p), Ar$_2^*$ |

**Table 2.** Gas-phase reactions utilized in the atmospheric pressure Ar-N$_2$-H$_2$ plasma kinetic model in addition to the gas-phase and surface reactions given in ref [1].

| No. | Reaction[a] | Rate coefficient (cm$^3$ s$^{-1}$)[b,c] | Reference |
|---|---|---|---|
| | **Electron processes** | | |
| R1 | $e$ + Ar → $e$ + $e$ + Ar$^+$ | BOLSIG+ | 33 |
| R2 | $e$ + Ar → $e$ + Ar(meta) | $5.0\times10^{-9} \times \exp(-12.64/T_e)$ | 31 |
| R3 | $e$ + Ar → Ar(r) + $e$ | $1.9\times10^{-9} \times \exp(-12.6/T_e)$ | 35 |
| R4 | $e$ + Ar → $e$ + Ar(4p) | $2.1\times10^{-8} \times \exp(-13.13/T_e)$ | 35 |
| R5 | $e$ + Ar$^+$ → Ar(4p) | $4\times10^{-13} \times T_e^{-0.5}$ | 36 |
| R6 | $e$ + $e$ + Ar$^+$ → Ar(4p) + $e$ | $5\times10^{-27} \times T_e^{-4.5}$ | 36 |
| R7 | $e$ + Ar(meta) → Ar$^+$ + $e$ + $e$ | $1.0\times10^{-7} \times \exp(-4.2/T_e)$ | 35 |
| R8 | $e$ + Ar(r) → Ar$^+$ + $e$ + $e$ | $1.0\times10^{-7} \times \exp(-4.2/T_e)$ | 35 |
| R9 | $e$ + Ar(4p) → Ar$^+$ + $e$ + $e$ | $1.8\times10^{-7} \times T_e^{0.61} \times \exp(-2.61/T_e)$ | 35 |
| R10 | $e$ + Ar(meta) → Ar(r) + $e$ | $3.7\times10^{-7}$ | 35 |
| R11 | $e$ + Ar(meta) → Ar(4p) + $e$ | $8.9\times10^{-7} \times T_e^{0.51} \times \exp(-1.59/T_e)$ | 35 |
| R12 | $e$ + Ar(r) → Ar(meta) + $e$ | $9.1\times10^{-7}$ | 35 |
| R13 | $e$ + Ar(r) → Ar(4p) + $e$ | $8.9\times10^{-7} \times T_e^{0.51} \times \exp(-1.59/T_e)$ | 35 |
| R14 | $e$ + Ar$_2^*$ → Ar$_2^+$ + $e$ + $e$ | $9.0\times10^{-8} \times T_e^{0.7} \times \exp(-3.66/T_e)$ | 36 |
| R15 | $e$ + Ar$_2^*$ → Ar + Ar + $e$ | $1.0\times10^{-7}$ | 36 |
| R16 | $e$ + Ar$_2^+$ → Ar(4p) + Ar | $5.4\times10^{-8} \times T_e^{-0.66}$ | 36 |
| R17 | $e + ArH^+ \rightarrow Ar + H$ | $1.0 \times 10^{-10} \times (-1.5 + 14 \times T_e - 16 \times T_e^2 + 8.4 \times T_e^3 - 1.9 \times T_e^4 + 0.2 \times T_e^5 - 0.0082 \times T_e^6)$ | 37 |
| | **Radiative relaxation** | | |
| R18 | $Ar(r) \rightarrow Ar$ | $1.0 \times 10^5$ | 31 |
| R19 | $Ar(4p) \rightarrow Ar(r)$ | $3 \times 10^7$ | 31 |
| R20 | $Ar(4p) \rightarrow Ar(meta)$ | $3 \times 10^7$ | 31 |
| | **Ar ion reactions** | | |
| R21 | $H_2^+ + Ar \rightarrow ArH^+ + H$ | $2.10 \times 10^{-9}$ | 31 |
| R22 | $H_2^+ + Ar \rightarrow Ar^+ + H_2$ | $2.00 \times 10^{-10}$ | 31 |
| R23 | $H_3^+ + Ar \rightarrow ArH^+ + H_2$ | $3.70 \times 10^{-10}$ | 31 |
| R24 | $Ar^+ + H_2 \rightarrow ArH^+ + H$ | $8.7 \times 10^{-10}$ | 31 |
| R25 | $Ar^+ + H_2 \rightarrow H_2^+ + Ar$ | $1.8 \times 10^{-11}$ | 31 |
| R26 | $Ar^+ + H_2 \rightarrow Ar + H^+ + H$ | $1.0 \times 10^{-9}$ | 36 |
| R27 | $Ar^+ + H \rightarrow H^+ + Ar$ | $1.0 \times 10^{-10}$ | 36 |
| R28 | $Ar^+ + N_2 \rightarrow N_2^+ + Ar$ | $4.45 \times 10^{-10}$ | 35 |
| R29 | $Ar^+ + N_2 \rightarrow Ar + N^+ + N$ | $5.0 \times 10^{-12}$ | 36 |
| R30 | $Ar^+ + N \rightarrow Ar + N^+$ | $4.45 \times 10^{-10}$ | 27 |
| R31 | $Ar^+ + NH_2 \rightarrow NH^+ + H + Ar$ | $5.5 \times 10^{-11}$ | 36 |
| R32 | $Ar^+ + NH_3 \rightarrow NH_3^+ + Ar$ | $1.84 \times 10^{-9} \times 0.92$ | 38 |
| R33 | $Ar^+ + NH_3 \rightarrow NH_2^+ + H + Ar$ | $1.84 \times 10^{-9} \times 0.03$ | 38 |
| R34 | $Ar^+ + NH_3 \rightarrow ArH^+ + NH_2$ | $1.84 \times 10^{-9} \times 0.05$ | 38 |
| R35 | $Ar^+ + Ar \rightarrow Ar + Ar^+$ | $4.6 \times 10^{-10}$ | 36 |
| R36 | $N^+ + Ar \rightarrow N + Ar^+$ | $6.0 \times 10^{-14}$ | 27 |
| R37 | $N_2^+ + Ar \rightarrow Ar^+ + N_2$ | $2.81 \times 10^{-10}$ | 27 |
| R38 | $N_3^+ + Ar \rightarrow N_2 + N + Ar^+$ | $6.6 \times 10^{-11}$ | 27 |
| R39 | $N_4^+ + Ar \rightarrow Ar^+ + N_2 + N_2$ | $1.0 \times 10^{-11}$ | 27 |
| R40 | $Ar_2^+ + N_2 \rightarrow Ar^+ + Ar + N_2$ | $2.50 \times 10^{-10}$ | 38 |
| R41 | $Ar_2^+ + H_2 \rightarrow ArH^+ + Ar + H$ | $4.7 \times 10^{-10}$ | 38 |
| R42 | $Ar_2^+ + H \rightarrow H^+ + Ar + Ar$ | $5.0 \times 10^{-11}$ | 36 |
| R43 | $ArH^+ + H_2 \rightarrow H_3^+ + Ar$ | $6.3 \times 10^{-10}$ | 31 |
| R44 | $ArH_2^+ + H_2 \rightarrow ArH_3^+ + H$ | $1.19 \times 10^{-9}$ | 38 |
| R45 | $Ar_2H^+ + H_2 \rightarrow ArH_3^+ + Ar$ | $1.20 \times 10^{-10}$ | 38 |
| R46 | $ArH^+ + NH_3 \rightarrow NH_3^+ + H + Ar$ | $5.3 \times 10^{-10}$ | 31 |
| R47 | $ArH^+ + NH_3 \rightarrow NH_4^+ + Ar$ | $1.6 \times 10^{-9}$ | 31 |
| R48 | $ArH^+ + N_2 \rightarrow N_2H^+ + Ar$ | $8.0 \times 10^{-10}$ | 31 |
| R49 | $ArH_3^+ + N_2 \rightarrow N_2H^+ + Ar + H_2$ | $9.0 \times 10^{-10}$ | 38 |
| R50 | $ArH_3^+ + NH_3 \rightarrow NH_4^+ + Ar + H_2$ | $2.56 \times 10^{-9}$ | 38 |
| R51 | $H^- + Ar_2^+ \rightarrow H + Ar + Ar$ | $2.0 \times 10^{-7}$ | 36 |
| R52 | $H^- + ArH^+ \rightarrow H + H + Ar$ | $2.0 \times 10^{-7}$ | 36 |
| **Excited Ar and neutral reactions** | | | |
| R53 | $Ar(meta) + Ar(meta) \rightarrow Ar^+ + Ar + e$ | $6.4 \times 10^{-10}$ | 35 |
| R54 | $Ar(meta) + Ar(4p) \rightarrow Ar^+ + Ar + e$ | $1.0 \times 10^{-9}$ | 36 |
| R55 | $Ar(meta) + Ar(r) \rightarrow Ar^+ + Ar + e$ | $2.1 \times 10^{-9}$ | 35 |
| R56 | $Ar(meta) + N_2 \rightarrow N + N + Ar$ | $1.6 \times 10^{-11}$ | 35 |
| R57 | $Ar(meta) + H_2 \rightarrow H + H + Ar$ | $1.1 \times 10^{-10}$ | 31 |
| R58 | $Ar(r) + N_2 \rightarrow Ar + N + N$ | $1.6 \times 10^{-11}$ | 35 |
| R59 | $Ar(r) + H_2 \rightarrow Ar + H + H$ | $1.1 \times 10^{-10}$ | 31 |
| R60 | $Ar(r) + Ar(4p) \rightarrow Ar^+ + Ar + e$ | $1.0 \times 10^{-9}$ | 36 |
| R61 | $Ar(4p) + Ar(4p) \rightarrow Ar^+ + Ar + e$ | $1.0 \times 10^{-9}$ | 36 |
| R62 | $Ar_2^* + H_2 \rightarrow H + H + Ar + Ar$ | $7.6 \times 10^{-11}$ | 39 |
| R63 | $Ar_2^* + Ar_2^* \rightarrow Ar_2^+ + Ar + Ar + e$ | $5.0 \times 10^{-10}$ | 36 |
| R64 | $Ar_2^* \rightarrow Ar + Ar$ | $6.0 \times 10^{7}$ | 36 |
| R65 | $Ar + N_2 \rightarrow N + N + Ar$ | $4.3 \times 10^{-10} \times \exp(-86460/T_g)$ | 36 |
| R66 | $N_2(B3) + Ar \rightarrow N_2(A3) + Ar$ | $3.0 \times 10^{-13}$ | 35 |
| R67 | $N_2(a'1) + Ar \rightarrow N_2(B3) + Ar$ | $1.0 \times 10^{-14}$ | 35 |
| **Penning excitation** | | | |
| R68 | $Ar(meta) + N_2 \rightarrow N_2(C3) + Ar$ | $3.0 \times 10^{-11}$ | 30 |
| R69 | $Ar(meta) + N_2 \rightarrow N_2(B3) + Ar$ | $9.8 \times 10^{-12}$ | 30 |
| R70 | $Ar(r) + N_2 \rightarrow N_2(C3) + Ar$ | $3.0 \times 10^{-11}$ | 30 |
| R71 | $Ar(r) + N_2 \rightarrow N_2(B3) + Ar$ | $9.8 \times 10^{-12}$ | 30 |
| **Recombination** | | | |
| R72 | $N + N + Ar \rightarrow N_2 + Ar$ | $2.3 \times 10^{-32}$ | 36 |
| R73 | $H + H + Ar \rightarrow H_2 + Ar$ | $6.4 \times 10^{-33} \times (300/T_g)$ | 36 |
| R74 | $N^+ + N + Ar \rightarrow N_2^+ + Ar$ | $1.0 \times 10^{-29}$ | 35 |
| R75-77 | $Ar + Ar + M \rightarrow Ar_2^* + Ar$ | $1.1 \times 10^{-32}$ | 36 |
| R78 | $N + N + Ar \rightarrow N_2(B3) + Ar$ | $(2.0/6.5) \times 8.27 \times 10^{-34} \times \exp(500/T_g)$ | 35 |
| R79 | $H + H + NH_3 \rightarrow H_2 + NH_3$ | $1.4 \times 10^{-31}$ | 36 |

**Reactions with NH species**

| | | | |
|---|---|---|---|
| R80-82 | $NH_3 + M \rightarrow Ar + NH_3^+ + e$ | $4.2 \times 10^{-11}$ | 36 |
| R83-85 | $NH_3 + M \rightarrow Ar + NH_2 + H$ | $5.8 \times 10^{-11}$ | 36 |
| R86-88 | $NH_3 + M \rightarrow Ar + NH + H + H$ | $5.2 \times 10^{-11}$ | 36 |
| R89-91 | $NH_3 + M \rightarrow Ar + NH + H_2$ | $5.8 \times 10^{-12}$ | 36 |
| R92 | $Ar + NH_3 \rightarrow H_2 + NH + Ar$ | $1.1 \times 10^{-9} \times \exp(-47032/T_g)$ | 36 |
| R93 | $Ar + NH_2 \rightarrow H + NH + Ar$ | $2.2 \times 10^{-9} \times \exp(-38224/T_g)$ | 36 |
| R94 | $Ar + NH \rightarrow H + N + Ar$ | $3.0 \times 10^{-10} \times \exp(-37615/T_g)$ | 36 |

[a]Rate coefficients for three body reactions have units of cm$^{-6}$ s$^{-1}$. For radiation reactions, the units are s$^{-1}$.
[b]M represents Ar(meta), Ar(r), and Ar(4p).
[c]$T_e$ is in eV and $T_g$ is in K unless otherwise specified.

Another addition to our previous CSTR ZDPlasKin model[15] is the ambipolar diffusion process to account for the loss of positive ions to the wall, which is described in general as

$$M^+ \rightarrow M^+(w) \qquad (5)$$

in ZDPlasKin. The rate coefficients were calculated by[30]

$$r = \frac{A}{V}\sqrt{\frac{kT_e}{m_i}}, \qquad (6)$$

where $m_i$ is the species mass, $T_e$ is the electron temperature, $k$ is the Boltzmann constant, $V$ is the volume, and $A$ is the effective surface area. Eley-Rideal (E-R) surface reactions and radical adsorption at surfaces have rate coefficients calculated by

$$k = \left[\frac{\Lambda^2}{D} + \frac{V}{A}\frac{2(2-\gamma)}{\bar{v}\gamma}\right]^{-1} S_T^{-1}, \qquad (7)$$

where for the gas phase species, $\Lambda$ is the diffusion length, $D$ is the diffusion coefficient, $\gamma$ is the sticking coefficient, and $\bar{v}$ is the thermal velocity; reactor dimensions are defined by $V$ the discharge volume, *A* the surface area that the gas species interact with, and $S_T$ the total surface sites in number density. A more detailed discussion of these terms can be found in Chen et al.[15] Combining the discharge volume and effective surface area gives a characteristic length, or the volume-to-surface area ratio (V/A). We excluded Langmuir-Hinshelwood (L-H) reactions since the dissociative adsorption of vibrationally excited $N_2$ on quartz is unlikely.[16] Moreover, we previously found that E-R reactions occur faster than L-H reactions under typical DBD plasma conditions and that the conditions (low reduced electric field) under which L-H reactions occur faster than E-R reactions and radical adsorption are challenging for plasma generation.[15,16]

E-R reactions ("surf" denotes a surface site) in our model include

$$H + N(s) \rightarrow NH(s), \qquad (8)$$

$$H + NH(s) \rightarrow NH_2(s), \qquad (9)$$

$$NH + H(s) \rightarrow NH_2(s), \qquad (10)$$

$$H + NH_2(s) \rightarrow NH_3 + surf, \qquad (11)$$

$$NH_2 + H(s) \rightarrow NH_3 + surf, \qquad (12)$$

$$H_2 + NH(s) \rightarrow NH_3 + surf. \qquad (13)$$

We also included the radical adsorption reactions

$$N + surf \rightarrow N(s), \qquad (14)$$

$$H + surf \rightarrow H(s), \qquad (15)$$

$$NH + surf \rightarrow NH(s). \qquad (16)$$

$N_2$ conversion was calculated as

$$\text{N}_2 \text{ conversion} = \frac{n_{\text{NH}_3}}{2n_{\text{N}_2,0}}, \tag{17}$$

where $n_{\text{NH}_3}$ is the NH$_3$ number density in the effluent and $n_{\text{N}_2,0}$ is the N$_2$ number density in the input for the reactor.

## 3. Results

### 3.1. Benchmarking model results and sensitivity analysis

We calculated N$_2$ conversions for typical input parameters that are reported and characteristic of DBD plasmas used for plasma-assisted catalytic ammonia synthesis. We configured the simulation setup to replicate the coaxial DBD reactor conditions of Liu et al.[25] Unless otherwise specified, the reactor conditions are given by the reduced electric field (E/N) = 71 Td, Ar % = 30%, Q = 100 ml/min, V = 1.508 ml, and gas pressure ratio N$_2$:H$_2$ = 1:1. The total gas pressure (P) and gas temperature (T$_g$) are 760 Torr and 300 K, respectively. The specific input energy, 15 kJ L$^{-1}$, is defined as the ratio of energy dissipated per minute and gas flow rate. The electron density (n$_e$) was not reported in Liu et al.[25] We examined the range of n$_e$ from 10$^6$-10$^9$ cm$^{-3}$ that is typically found in packed-bed reactors[1,15,16] measured by voltage-charge Lissajous plots. Our group also observed n$_e$ values typically on the order of 10$^6$-10$^8$ cm$^{-3}$ in near-atmospheric pressure Ar-N$_2$-H$_2$ plasmas in an empty tube flow reactor using the Lissajous method. The reactor-dependent parameters of surface-to-area ratio (V/A) and diffusion length ($\Lambda$) are important to surface reactions. We used the volume-to-surface area ratio, V/A = 1×10$^{-6}$ m, estimated from the geometric area, 0.20 m$^2$/g, and the density of quartz wool, 2.2-2.6 g/cm$^3$,[40,41] for the reactor used in Liu et al.[25] Although the value of V/A from this estimation can be obscured depending on the effective surface area that the plasma species interact with, our previous modeling work found that N$_2$ conversion is not sensitive to V/A values once they reach the order of magnitude used for this simulation.[15] The diffusion length ($\Lambda$) is challenging to capture because it is also affected by the macroscopic geometrical configuration of the wool fibers. In practice, the space gaps between fibers of quartz wool typically range from 1 to 100 um.[42,43]

Fig. 1 shows the calculated N$_2$ conversions for different electron densities and diffusion lengths. As the electron density increases, we observe increasing N$_2$ conversion. Liu et al. measured N$_2$ conversion of 0.18% for an input energy of 15 kJ L$^{-1}$.[25] At $n_e = 1.8 \times 10^8$ cm$^{-3}$, the calculated N$_2$ conversion matched well their experimental result. The calculated results using three diffusion lengths vary by less than 0.05%. A more detailed sensitivity analysis for the effect of diffusion length on N$_2$ conversion is illustrated in Fig. 2. The choice of diffusion length does not significantly affect the calculated N$_2$ conversion over this range and the calculated N$_2$ conversions vary less than 0.05%.

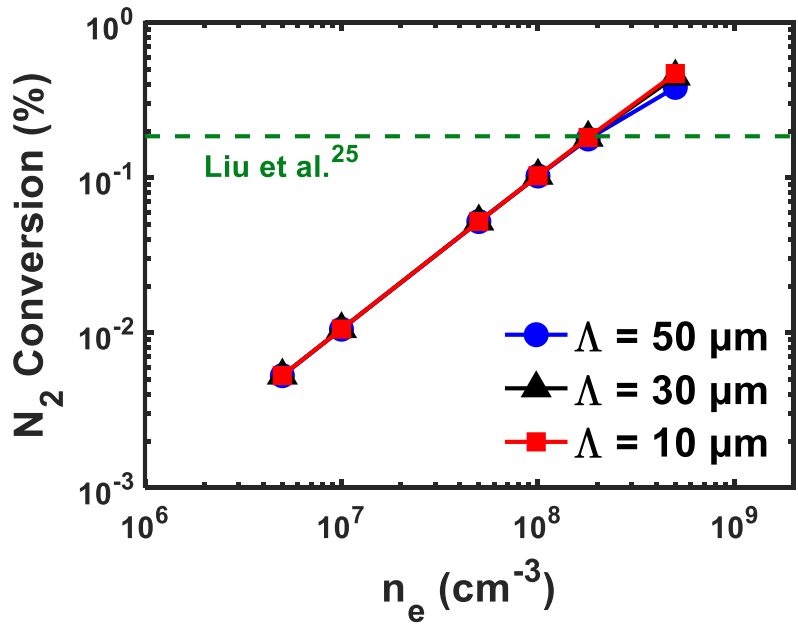

**Figure 1.** Calculated N$_2$ conversion as a function of electron density for different diffusion lengths. The dashed line denotes an experimental value from Liu et al.[25]

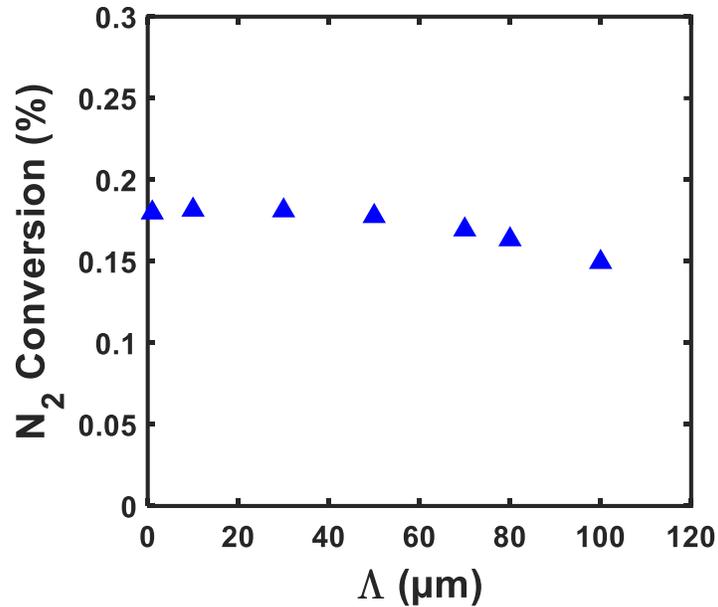

**Figure 2.** Effect of diffusion length on calculated $N_2$ conversion. Plasma conditions: E/N = 71 Td, $n_e$ = 1.8×10$^8$ cm$^{-3}$.

## 3.2. Reaction pathways of NH$_3$ synthesis for an atmospheric pressure Ar-N$_2$-H$_2$ plasma in a quartz wool-packed DBD reactor

The ZDPlasKin CSTR kinetic model constructed as described above was used to identify the major pathways for NH$_3$ production in an Ar-N$_2$-H$_2$ plasma in a quartz wool-packed DBD reactor under a specific condition of interest. Fig. 3 shows the production rates of major species and reaction pathways for NH$_3$ production assuming $n_e$ = 1.8×10$^8$ cm$^{-3}$ and $\Lambda$ = 10 µm (Sec. 3.1.). Reactions that were originally included in the ZDPlasKin model in refs. [1] and [15] and are important for NH$_3$ formation are listed in Table 3. The rates were normalized by the total formation rate of NH$_3$ via surface pathways, i.e., the sum of rates for R95-97, which are the final steps of producing NH$_3$. For the dissociation of N$_2$ and H$_2$, the relative production rate shown in Fig. 3 is double the relative steady-state reaction rate since two N and two H radicals are produced per dissociation reaction of N$_2$ and H$_2$, respectively. E-R reactions account for nearly all of the NH$_3$ production, where the most dominant is H$_2$ + NH(s) → NH$_3$ + surf (R97, 41% of NH$_3$ formation), followed by H + NH$_2$(s) → NH$_3$ + surf (R95, 32%) and NH$_2$ + H(s) → NH$_3$ + surf (R96, 27%). NH$_3$ produced from gas-phase radical reactions, i.e., not involving a surface, account for less than 0.4% of the total NH$_3$ formation.

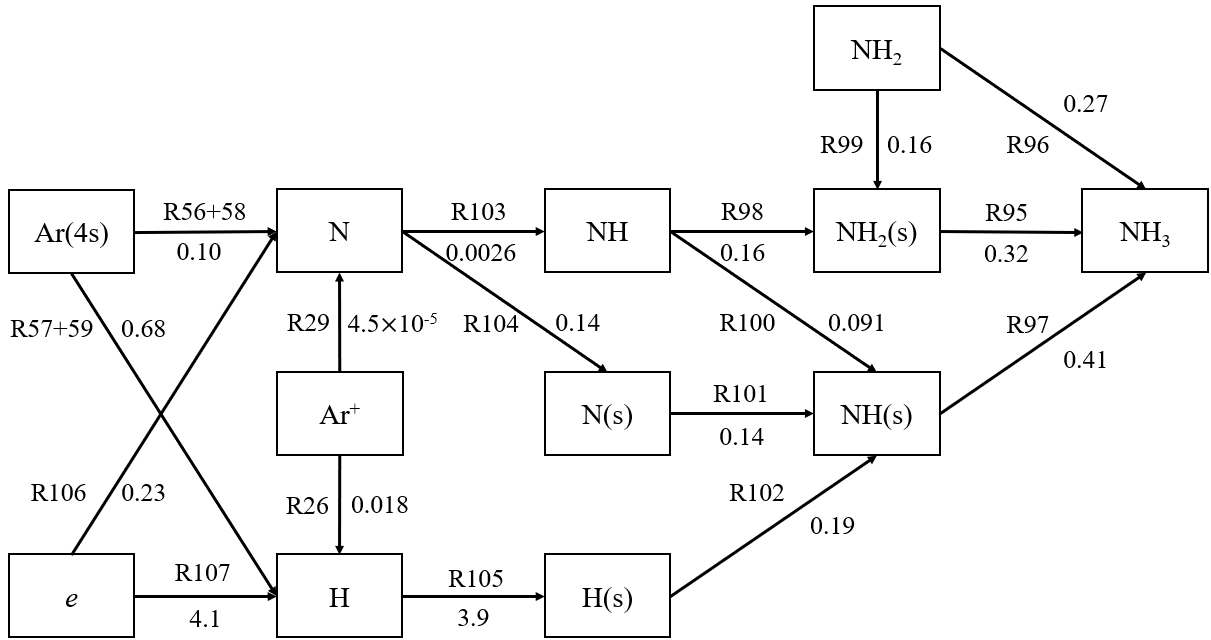

**Figure 3.** Reaction network for plasma-assisted catalytic NH$_3$ synthesis in a quartz wool-packed DBD reactor. Simulation conditions: $n_e$ = 1.8×10$^8$ cm$^{-3}$, E/N = 71 Td, $\Lambda$ = 10 µm. Production rates were normalized by the total formation rate of NH$_3$ via the E-R surface reaction pathways (R95-97) in Table 3.

**Table 3.** Major reactions for $NH_3$ formation involving N and H species identified for an atmospheric pressure Ar-$N_2$-$H_2$ DBD plasma at the conditions of E/N = 71 Td and $n_e$ = $1.8 \times 10^8$ cm$^{-3}$.

| No. | Reaction |
|---|---|
| R95 | $H + NH_2(s) \rightarrow NH_3 + surf$ |
| R96 | $NH_2 + H(s) \rightarrow NH_3 + surf$ |
| R97 | $H_2 + NH(s) \rightarrow NH_3 + surf$ |
| R98 | $NH + H(s) \rightarrow NH_2(s)$ |
| R99 | $NH_2 + surf \rightarrow NH_2(s)$ |
| R100 | $NH + surf \rightarrow NH(s)$ |
| R101 | $H + N(s) \rightarrow NH(s)$ |
| R102 | $N + H(s) \rightarrow NH(s)$ |
| R103 | $N + H_2^* \rightarrow NH + H$ |
| R104 | $N + surf \rightarrow N(s)$ |
| R105 | $H + surf \rightarrow H(s)$ |
| R106 | $e + N_2 \rightarrow e + 2N$ |
| R107 | $e + H_2 \rightarrow e + 2H$ |

In the steady state, the production processes of NH(s) rely mainly on gas-phase N and H radicals and adsorbed N and H adatoms (surface-bound atoms). The direct adsorption of gas-phase NH radicals (R100) contributes 22% compared to 33% via the E-R reaction of $H + N(s) \rightarrow NH(s)$ (R101) and 45% via the E-R reaction of $N + H(s) \rightarrow NH(s)$ (R102). In the steady state, the dominant route of gas-phase NH radical production is the electron-impact dissociation of $NH_3$. We note that the reaction between gas-phase N radicals and electronically excited $H_2$ molecules in the gas phase (R103) only has a relative NH production rate of 0.0026 at steady state. Hong et al. found that gas-phase NH radicals were important in the early phase of $NH_3$ formation through the three-body gas-phase reactions

$$NH + H_2 + H_2 \rightarrow NH_3 + H_2, \qquad (18)$$

$$NH + N_2 + H_2 \rightarrow NH_3 + N_2, \qquad (19)$$

which are more prominent at atmospheric pressure than in low-pressure regimes.[1] Gas-phase NH radicals were mainly produced by the electron-impact dissociation of $NH_3$ after the build-up of $NH_3$ in the gas phase, which allows electron-impact dissociation of $NH_3$ into gas-phase $NH_x$ radicals.

The formation of adsorbed NH(s) via R100-102 and NH radicals in the gas phase via R103 highlights the importance of both gas-phase N radicals and adsorbed N adatoms. We found that gas-phase N radical adsorption and the formation of adsorbed NH(s) by H + N(s) → NH(s) have the same relative production rates. This means that adsorbed N adatoms are not depleted by faster side reactions that do not lead to $NH_3$ formation and promptly contribute via R97 to the $NH_3$ formation reaction network. We found that excited argon species via R56+58 contribute to 28% of the total gas-phase N radical production, but direct dissociation of gas-phase $N_2$ molecules by collisions with electrons is the major source of N radicals in the gas phase comprising 65%. Furthermore, the contribution of excited argon species to gas-phase N radical production derives almost completely from Ar(meta) and Ar(r) (the 4s excited states). The relative production rate of gas-phase N radicals via R29 indicates that $Ar^+$ has a negligible contribution to the direct dissociation of $N_2$ in the gas phase. At a low degree of ionization, we expect radicals to play a more significant role than ions. For H radical production, $H_2$ dissociation by Ar species is only 17% of that caused by electron-impact dissociation. Thus, one potential cause of the slight enhancement of $N_2$ conversion observed in the experiments of Liu et al.[25] is the role of metastable Ar species in dissociating $N_2$ molecules.

## 4. Discussion

The zero-dimensional kinetic modeling of atmospheric pressure Ar-$N_2$-$H_2$ nonthermal plasma we carried out enables a detailed reaction analysis regarding the effects of Ar addition on ammonia formation in a DBD plasma. Unless otherwise specified, the input plasma parameters for the discussion below are the same as given herein in Sec. 3.2, i.e., E/N = 71 Td, $n_e$ = 1.8×10$^8$ cm$^{-3}$, Ar % = 30%, Q = 100 ml/min, V = 1.508 ml, $N_2$:$H_2$ = 1:1, P = 760 Torr, and $T_g$ = 300 K.

**4.1. NH$_3$ decomposition by argon species**

Energy transfer between excited Ar species and NH$_3$ in the gas phase may result in NH$_3$ decomposition. The 4s and 4p metastable Ar states have excitation energies of 11.6 and 13.1 eV, respectively, which are sufficient to cause NH$_3$ dissociation.[36] At steady state, NH$_3$ is predominantly decomposed in the plasma by electron impact with a rate of 7.1×10$^{16}$ cm$^{-3}$ s$^{-1}$, corresponding to 65% of the total NH$_3$ formation rate via E-R surface reactions. Argon species play a weaker role in NH$_3$ dissociation. Major argon species that participate in NH$_3$ decomposition reactions include Ar(meta), Ar(r), Ar(4p), Ar$^+$, and ArH$^+$. These reactions are described by R80-91 for Ar(meta), Ar(r), and Ar(4p), R32-34 for Ar$^+$, and R46-47 for ArH$^+$. The contribution from Ar(4p) is the most significant, with a total rate for NH$_3$ decomposition of 3.2×10$^{15}$ cm$^{-3}$ s$^{-1}$. However, this rate is still 95% slower than electron-impact dissociation of NH$_3$. Dissociation rates of NH$_3$ are one order of magnitude lower by Ar(meta) and Ar(r), and three orders of magnitude lower by Ar$^+$ and ArH$^+$, compared to that by Ar(4p). It is worth noting that dissociation by Ar(meta), Ar(r), Ar(4p), and Ar$^+$ can also directly lead to the production of the gas-phase radicals NH and NH$_2$, which can be utilized to produce NH$_2$(s) by radical adsorption. We thus conclude that argon does not have a profound effect on the destruction of NH$_3$ in the plasma.

**4.2. The role of Ar$^+$ ions and dissociative recombination of $N_2^+$ ions**

Gas-phase Ar and N$_2$ have similar first ionization energies (15.76 eV and 15.58 eV).[26] Thus, charge transfer between gas-phase Ar and N$_2$ happens readily in the plasma.[28] In steady state, we found that charge transfer from gas-phase $Ar^+$ to $N_2$ (eq. 20) happens four times as fast as the reverse process (eq. 21).

$$Ar^+ + N_2 \rightarrow Ar + N_2^+ \quad (20)$$

$$Ar + N_2^+ \rightarrow Ar^+ + N_2 \quad (21)$$

Nakajima et al. proposed that charge transfer reactions enhance gas-phase N radical density in the plasma by the subsequent dissociative recombination of $N_2^+$.[24] Kang et al. found that the dissociative recombination

$$e + N_2^+ \rightarrow N + N \quad (22)$$

dominates N radical production pathways when Ar is the main discharge gas component in a 200-mTorr Ar-$N_2$ plasma.[28]

Under the conditions of the calculations herein, the dissociative recombination reactions of $N_2^+$ are ten orders of magnitude slower than that of electron-impact dissociation of $N_2$. We first attributed the observation to our low degree of ionization and an abundance of side reactions enabled by the highly collisional environment at atmospheric pressure. But Liang et al.[35] and Bogaerts[27] also found the production of gas-phase N radical by the dissociative recombination reactions of $N_2^+$ is insignificant at low pressure and higher electron density in Ar-$N_2$ plasma. In our model of an atmospheric pressure Ar-$N_2$-$H_2$ plasma, we found $N_2^+$ is mostly depleted by collision with $H_2$ molecules to produce gas-phase $N_2H^+$ rather than with neutralization by electrons or charge transfer with Ar, and the gas-phase $N_2H^+$ can subsequently play a minor role in consuming $NH_3$ via the gas-phase reaction $N_2H^+ + NH_3 \rightarrow NH_4^+ + N_2$.

### 4.3. Penning excitation by argon species

Penning excitation of $N_2$ in the gas phase by excited Ar species (R68-71) was investigated. The energy transferred from excited Ar species in turn dissipates via reactions and decay processes of electronically excited $N_2$ species. One such process is the radiative decay from the higher energy state $N_2(C3)$ to the lower energy state $N_2(A3)$ (eq. 23).

$$N_2(C3) \rightarrow \ldots \rightarrow N_2(A3) \tag{23}$$

More details regarding the interactions between electronically excited $N_2$ species can be found in Hong et al.[1] It has also been suggested that metastable-metastable pooling dissociation between two $N_2(A3)$ molecules can effectively generate N radicals in the gas phase via[28]

$$N_2(A3) + N_2(A3) \rightarrow N_2 + N + N. \tag{24}$$

However, we found that the rate of N radical production from $N_2(A3)$ dissociation is several orders of magnitude lower than $N_2$ dissociation by Ar(meta) and Ar(r) under the plasma conditions for ammonia synthesis in Fig. 3. We note that gas-phase Ar(meta) is mostly depleted by collisions with $H_2$ leading to dissociation (R57), with slower Ar(meta) depletion caused by Penning excitation of $N_2$ (R68+69) and collisions with $N_2$ leading to dissociation (R56). Among interactions between Ar(meta) and $N_2$, the steady state reaction rates for R56, R68, and R69 are $3.1 \times 10^{15}$ cm$^{-3}$ s$^{-1}$, $5.8 \times 10^{15}$ cm$^{-3}$ s$^{-1}$, and $1.9 \times 10^{15}$ cm$^{-3}$ s$^{-1}$, respectively. The utility of Ar(meta) for the Penning excitation of $N_2$ to produce $N_2(B3)$ and $N_2(C3)$ causes a 2.5 times higher rate of Ar(meta) depletion than collisions with $N_2$ leading to dissociation of $N_2$. Yet, the contribution to $N_2(C3)$ production by both Ar(meta) and Ar(r) (eq. 25) is only 8.0% of that caused by electron-impact excitation (eq. 26). For $N_2(B3)$ formation, the contribution by both Ar(meta) and Ar(r) declines to 1.2% of that caused by electron-impact excitation. The high Penning excitation reaction rates suggest that Penning excitation is a pronounced mechanism for quenching Ar(meta) by $N_2$, but a minor source of electronically excited $N_2$ under these conditions.

$$\text{Ar(meta), Ar(r)} + N_2 \rightarrow \text{Ar} + N_2(C3) \qquad (25)$$

$$e + N_2 \rightarrow e + N_2(C3) \qquad (26)$$

**4.4. $N_2$ conversion and Ar contribution as affected by electron density and reduced electric field**

Fig. 4 shows $N_2$ conversion as a function of the reduced electric field (E/N) and electron density ($n_e$). These calculated $N_2$ conversions are intended to provide guidance for the different plasma conditions needed to determine the optimal operating conditions. $N_2$ conversion increases with both E/N and $n_e$, which is expected given the higher input power. Within this range of E/N and $n_e$ values, E-R surface reactions dominate $NH_3$ formation. The sources of N radicals in the gas phase are dominated by $N_2$ dissociation by electron impact and collisions with Ar(meta) and Ar(r). We define the Ar contributing factor as the ratio of the rate of $N_2$ dissociation by both Ar(meta) and Ar(r) to that by electron-impact dissociation (eq. 27).

$$f_{Ar \rightarrow N} = \frac{r_{R56} + r_{R58}}{r_{R106}} \qquad (27)$$

Fig. 5 gives the Ar contributing factor in atmospheric pressure $Ar$-$N_2$-$H_2$ plasma for the various conditions we simulated. The Ar contributing factor never exceeds 1, indicating that electron-impact dissociation of $N_2$ always dominates. In cases when Ar is the dominant gas component in the plasma, $N_2$ dissociation by excited Ar has been found to surpass electron impact dissociation.[27,35] Our model predicts that the Ar contributing factor is maximized at E/N = 80 Td, when the mean electron temperature is 2.7 eV, and decreases with high $n_e$.

It is commonly assumed in the literature that the plasma is a source of vibrationally excited $N_2$ molecules, which can dissociatively adsorb at the catalytic surface via a lower energy barrier, and this enhances the rate of $N_2$ dissociation, which is the rate-determining step.[5,6,44] However, due to the reactivity of N radicals, they can outcompete excited $N_2$ molecules as a source of N(s) on a catalyst.[7] Engelmann et al. showed that reactions of plasma-generated radicals have more impact on $NH_3$ formation than catalytic dissociation of excited $N_2$ molecules in an atmospheric pressure $N_2$-$H_2$ DBD discharge at 400 K and E/N = 20-100 Td.[45] Although our model shows that Ar(meta) and Ar(r) in the plasma can potentially help promote the generation of gas-phase N radicals and enhance the subsequent E-R reactions to form $NH_3$, dissociation of $N_2$ molecules in the plasma is not catalyzed and inherently more energetically expensive than excitation. Given that electron-impact dissociation of gas-phase $N_2$ always dominates the production of gas-phase N radicals, the opportunity for diluting $N_2$-$H_2$ plasma with Ar may instead be found in plasma generation and production of excited gas-phase $N_2$ molecules rather than $N_2$ dissociation in the plasma. Our model suggests that $N_2$ molecules can effectively quench Ar(4s) via Penning excitation reactions and produce electronically excited gas-phase $N_2$ at atmospheric pressure and near room temperature. Akin to vibrationally excited $N_2$, dissociative adsorption of electronically excited $N_2$ may also occur on catalytic sites via a lower energy barrier.[1] Moreover, Ar dilution is known for enhancing the electron density in the plasma and sustaining the host plasma at lower reduced electric fields.[17,21,25,28] Lower values of E/N (1-30

Td) are also more favorable for the generation of vibrationally excited species compared to that in high E/N regimes.[46] Therefore, the utilization of Ar to tailor discharge characteristics and produce electronically excited gas-phase $N_2$ molecules under other plasma conditions are directions that should be investigated. Future studies may also examine conditions in which L-H reactions can compete with E-R reactions over a metal catalyst when Ar is added to $N_2$-$H_2$ plasmas.

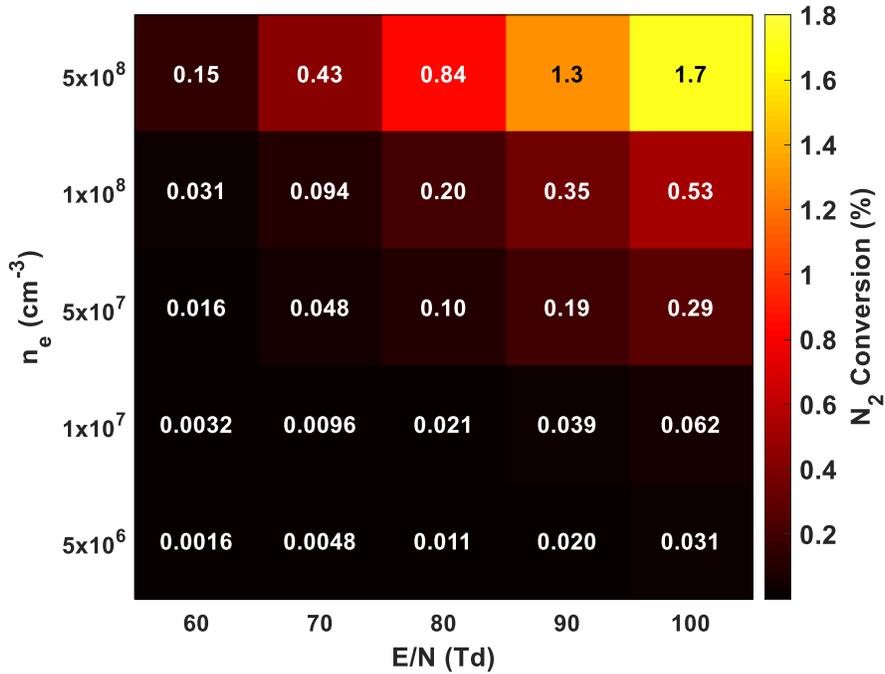

**Figure 4.** Dependence of $N_2$ conversion on the reduced electric field and electron density in a 1:1 $N_2$-$H_2$ plasma with 30% Ar in a quartz wool-packed DBD reactor at atmospheric pressure and near room temperature.

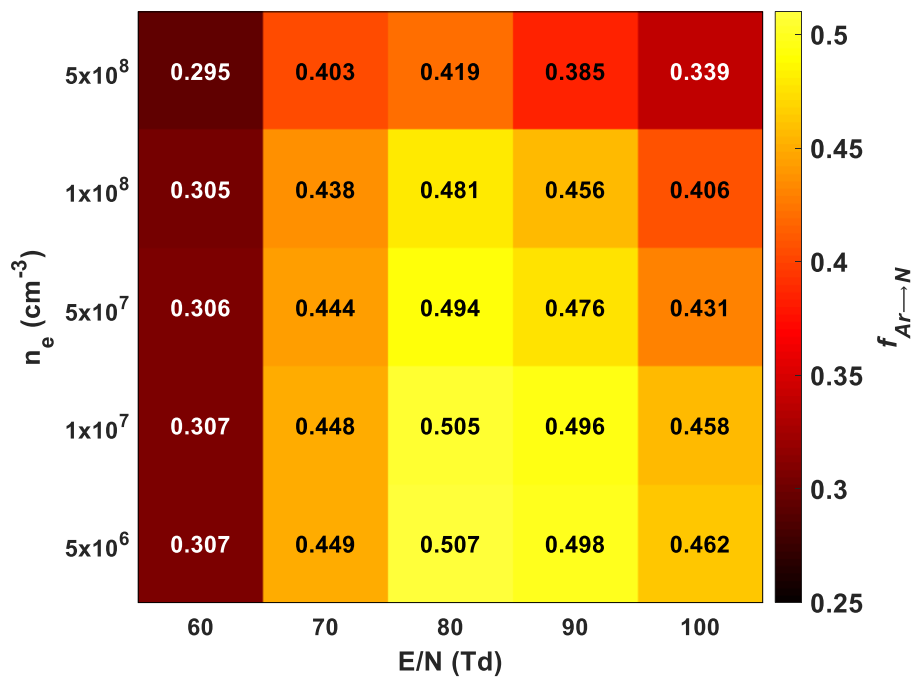

**Figure 5.** Dependence of the Ar contributing factor on the reduced electric field and electron density in a 1:1 $N_2$-$H_2$ plasma with 30% Ar in a quartz wool-packed DBD reactor at atmospheric pressure and near room temperature. The Ar contributing factor is defined as the ratio of the rate of $N_2$ dissociation by Ar(meta) and Ar(r) to that by electron-impact dissociation.

## 5. Conclusion

We have investigated the chemical pathways in atmospheric pressure Ar-$N_2$-$H_2$ nonthermal plasmas for ammonia synthesis by using a 0D nonthermal plasma kinetic model adapted to a coaxial DBD flow reactor packed with quartz wool. By comparing the modeling results with existing experimental results, we identified the target conditions for the analysis of the $NH_3$ synthesis pathway. It was observed that the model is not sensitive to the choice of diffusion length in the regime that we investigated, which is an important parameter for calculating reaction rates for processes involving the surfaces in the reactor. Upon diluting a 1:1 $N_2$-$H_2$ plasma with 30% Ar, the kinetics of formation of gas-phase N and H radicals and surface-bound N and H adatoms are supplemented by reactions of gas-phase $N_2$ and $H_2$ with excited Ar species. Under the conditions that we identified, we found that Ar can contribute to 28% of N radical production in the gas phase, which subsequently produces gas-phase NH radicals and adsorbed $NH_x(s)$ species in the early steps of $NH_3$ synthesis. Adsorbed $NH_x(s)$ species then dominate the steady-state formation of $NH_3$ via Eley-Rideal (E-R) reactions. Regarding the role of Ar in increasing the $NH_3$ synthesis rate, analysis of the modeling results reveals that the Ar contributions originate from excited Ar species in the 4s state. Moreover, charge transfer reactions between $Ar^+$ and $N_2$, when followed by dissociative recombination of $N_2^+$, are not a significant source of gas-phase N radical production under the given conditions, which is dominated by electron-impact dissociation of $N_2$. Thus the model predicts that argon is unlikely to have a dominant effect on the efficiency of $NH_3$ formation via a chemical pathway due to the dominant role of electrons in this process under most conditions. Instead, experimentally observed enhancement in $NH_3$ production upon Ar dilution may be due to both changes to the chemistry and the electron number density and reduced electric field in the discharge. The utilization of Ar to optimize plasma discharge conditions and produce electronically excited gas-phase $N_2$ molecules under the presence of catalysts are future subjects of investigation.

**Acknowledgment**

This material is based upon work supported by the U.S. Department of Energy, Office of Science, Office of Fusion Energy Sciences under award number DE-SC0020233.